# Engineered substrates for domain control in CrSe thin-film growth: Single-domain formation on lattice-matched YSZ(111) substrate

Yusuke Tajima[1], Junichi Shiogai[1,2,a)], Masayuki Ochi[1,3],

Kazutaka Kudo[1,2], and Jobu Matsuno[1,2]

[1]*Department of Physics, Osaka University, Toyonaka, Osaka 560-0043, Japan*

[2]*Division of Spintronics Research Network, Institute for Open and Transdisciplinary Research Initiatives, Osaka University, Suita, Osaka 565-0871, Japan*

[3]*Forefront Research Center, Osaka University, Toyonaka, Osaka 560-0043, Japan*

[a)] Author to whom correspondence should be addressed.
Electronic mail: junichi.shiogai.sci@osaka-u.ac.jp

**ABSTRACT**

Epitaxial thin-film growth is a versatile and powerful technique for achieving a precise control of composition, stabilizing non-equilibrium phases, tailoring growth orientation, as well as forming heterointerfaces of various quantum materials. For synthesis of highly crystalline thin films, in-depth understanding of epitaxial relationship between the desired thin film and the single-crystalline substrates is necessary. In this study, we investigate epitaxial relationship in thin-film growth of triangular-lattice antiferromagnet CrSe on the (001) plane of $Al_2O_3$ and the lattice-matched (111) plane of yttria-stabilized zirconia (YSZ) substrates. Structural characterization using out-of-plane and in-plane x-ray diffraction shows that the presence of 19.1°-twisted domains of CrSe significantly dominates the aligned domain on the $Al_2O_3$ substrate while it reveals a single-domain formation on the YSZ substrate. The stability of the 19.1°-twisted domain rather than the aligned domain can be explained by rotational commensurate epitaxy, which is well reproduced by density functional theory calculations. The single-domain CrSe thin film on the YSZ substrate exhibits a superior metallic conductivity compared to the twisted-domain thin film on the $Al_2O_3$ substrate, implying contribution of the grain boundary scattering mechanism to electrical transport.

# I. INTRODUCTION

Two-dimensional (2D) network of edge-shared octahedra of chromium chalcogenides and pnictides Cr-$X$ ($X$ = S, Se, Te, and Sb) constitutes various magnetic materials by selecting chalcogen and pnictogen elements [1,2] or intercalated ions [3]. The simplest compounds are NiAs-type (space group: $P6_3/mmc$) Cr$X$ binary systems [Fig. 1(a)], which show a rich variety of magnetic states depending on $X$ elements among room-temperature ferromagnet in CrTe [4,5], triangular-lattice antiferromagnet with non-coplanar spin structure in CrSe [6,7], as well as high-temperature altermagnet in CrSb [8]. Further, NiAs derivative structures of Cr-$X$ binary systems contain self-intercalated Cr atoms between the Cr$X_2$ networks [9,10]. In Cr$_{1+x}$Te$_2$ ($x$ = 0, 1/4, 1/3, 1/2, 1) compounds, the occupancy of Cr intercalants and their degree of order have a large influence on magnetic anisotropy, magnetic transition temperature, and Fermi level [1,4,11]. The Cr$X_2$ 2D network can also be a building block of ternary magnetic compounds $A$Cr$X_2$ (space group: $R3m$ or $R\bar{3}m$), where $A^+$ ions are intercalated between the stacked Cr$X_2$ networks. This class of materials has attracted significant interest in the viewpoints of energy harvesting materials by a superior ionic conductivity and ultra-low thermal conductivity [12]. They are also promising spintronic materials owing to their strong spin-orbit interaction both at surface [13] and bulk [14] as well as multiferroic properties [15]. With the aim of further functionalizing their intriguing properties of Cr$X$-based magnetic compounds, thin-film growth is advantageous in many aspects such as possible tunability of carrier density with electric gating in semiconducting [16] and semimetallic [17] states, and that of relative contribution from surface state [13] or interfacial effect [18,19,20] to the electrical transport properties.

For the epitaxial thin-film growth of materials with rhombohedral or hexagonal lattices, the (001) plane of sapphire Al$_2$O$_3$ or the (111) plane of cubic crystal are commonly used as single crystalline substrates. Although the Al$_2$O$_3$(001) substrate offers an atomically flat surface [21], the in-plane lattice constant is $a_{Al_2O_3}$ = 4.7589 Å, being much larger than many Cr$X$-based compounds. In previous works, successful growth of the fully aligned and single-domain CrTe films has been reported using the Al$_2$O$_3$(001) substrate with epitaxial relationship of CrTe(001)[100]//Al$_2$O$_3$(001)[100] (hexagonal Miller indices are used, hereafter) [4,5], despite the large lattice mismatch ($a_{film}$ − $a_{substrate}$)/$a_{substrate}$ = −16.4% between CrTe ($a_{CrTe}$ = 3.978 Å) and Al$_2$O$_3$. This is probably because the weak Cr-Te bonding, as suggested by a small exfoliation energy of the (002) face of CrTe (Ref. [22]), allows van der Waals (vdW) epitaxy-like growth mode [23] even though the NiAs-type Cr$X$ is not technically categorized as vdW materials. Recently, we have demonstrated $c$-axis oriented thin-film growth of NiAs-type CrSe on the Al$_2$O$_3$(001) substrate by a precise tuning of stoichiometry, which allows observation of metallic conductivity with unique magneto-transport properties originating from triangular-lattice antiferromagnet [7]. The in-plane lattice constant of CrSe is $a_{CrSe}$ = 3.684 Å [2], which corresponds to –22.6% mismatch with the $Al_2O_3$(001) substrate being much larger than the CrTe case, as summarized in Fig. 1(b). For developing thin-film devices based on CrSe and its related compounds, in-depth understanding of epitaxy and reduction of defects of the base octahedron networks are subject of interest.

In this study, we have identified the in-plane orientation of the CrSe thin films grown on the $Al_2O_3$(001) and yttria-stabilized zirconia (YSZ) (111) substrates. We employ the cubic YSZ(111) substrate ($a_{YSZ}$ = 5.12 Å) as a lattice-matched substrate. The nearest O-O distance on the (111) plane of YSZ is $a_{YSZ}/\sqrt{2}$ = 3.62 Å, which is close to $a_{CrSe}$ with a small lattice mismatch of −1.77%, as shown in Fig. 1(b). On both substrates, we have found the CrSe thin films are grown along the *c*-axis direction with comparable *c*-axis lengths. In contrast, we have observed a distinct difference in in-plane crystallographic orientation. For the $Al_2O_3$(001) substrate, the strong XRD peaks originating from the in-plane ±19.1°-twisted domains are observed, where the [100] direction of CrSe is rotated by either +19.1° or −19.1° from the [100] direction of $Al_2O_3$, in addition to the rather weak peaks from the aligned domain. We ascribe this specific angle to the rotational commensurate atomic arrangement of the 3×4 unit-cells (UC) of CrSe on the 2×2 UC of Al$_2$O$_3$. This rotational commensurate epitaxy is further supported by the density functional theory (DFT) calculations showing that the twisted domains are energetically favorable rather than the aligned domains. On the other hand, the twisted-domain formation is suppressed on the lattice-matched YSZ(111) substrate. The CrSe thin film with the single domain on the YSZ substrate exhibits a superior conductivity to that with the multiple twisted domains on the Al$_2$O$_3$ substrate, particularly below magnetic transition temperature, implying that the finite contribution of grain boundary scattering to electrical transport.

## II. EXPERIMENTAL METHODS

The 50-nm-thick CrSe thin films were deposited on the Al$_2$O$_3$(001) and YSZ(111) substrates by pulsed-laser deposition (PLD) at the substrate temperature of 750 ºC and base pressure of about $7.5 \times 10^{-6}$ Torr, followed by in-situ deposition of the 10-nm-thick Se cap layer at room temperature. The PLD targets were synthesized by pelletizing commercially-available stoichiometric CrSe and Se powders (Kojundo Chemical Laboratory Co. Ltd) for deposition of the CrSe thin films and Se cap layer, respectively. The growth temperature of $T_{sub}$ = 750 ºC was optimized in previous study to obtain the stoichiometric composition ratio Cr/Se = 1 [7]. The structural characterization was performed by specular (out-of-plane) and off-specular (in-plane) rotation x-ray diffraction (XRD) with Cu-$K\alpha$1 used as the x-ray source. For simulating interface reconstruction of CrSe on the Al$_2$O$_3$(001) substrate, we performed DFT calculations for the slab model consisting of 4-layer CrSe and 6-layer Al$_2$O$_3$ stacked along the $c$ axis with an around 15 Å vacuum region. We used the Perdew-Burke-Ernzerhof parameterization of the generalized gradient approximation [24] with the $+U$ correction in the simplified rotationally invariant approach introduced by Dudarev *et al*. [25] and the DFT+D3 (BJ) dispersion correction [26,27] as implemented in Vienna *Ab initio* Simulation Package [28,29,30,31,32]. We set the Hubbard correction $U_{\text{eff}} = U - J = 5$ eV for Cr-$d$ orbitals. We set the core states of each atomic potential used in the projector augmented wave method [33] as [Ne], [He], [Ar], [Ar]$3d^{10}$ for Al, O, Cr, Se, respectively. We fixed the in-plane lattice constants using the experimental $a$-axis length of Al$_2$O$_3$, $a_{\text{Al}_2\text{O}_3} = 4.7589$ Å, and only optimized the atomic coordinates in the cell until the Hellmann-Feynman force on every atom became less than 0.01 eV/Å. The electrical transport property was evaluated by temperature dependence of longitudinal resistivity with standard four-terminal configuration using a lock-in technique and a $^4$He variable temperature insert (Oxford Instruments, plc).

## III. RESULTS AND DISCUSSION

Figure 1(c) shows the out-of-plane XRD pattern of the CrSe thin film deposited on the Al$_2$O$_3$(001) substrate showing only CrSe(00$l$) diffraction peaks, indicating that the $c$ axis is oriented normal to the plane as previously reported [7]. Similarly to the Al$_2$O$_3$ case, we also observed the $c$-axis-oriented growth of CrSe on the YSZ(111) substrate as shown in Fig. 1(d). Note that the only (001) diffraction peak was clearly observed in the CrSe film on the YSZ(111) substrate while diffraction peaks of the CrSe(002) and (004) were not detected because the intense YSZ(111) and (222) diffraction peaks overlap the CrSe(002) and (004) peaks, respectively. Therefore, the $c$-axis lattice constant of CrSe is determined by (001) in this study. Figures 1(e) and 1(f) show the out-of-plane XRD pattern around the CrSe(001) diffraction peaks for the Al$_2$O$_3$(001) and YSZ(111) substrates. From the CrSe(001) diffraction peak, the $c$-axis length of CrSe on Al$_2$O$_3$(001) and YSZ(111) are estimated as 5.953 Å and 5.935 Å substrates, respectively. The difference in the $c$-axis length is not significant within the subtle variations among investigated samples and these values are consistent with our previous report [7].

To detect in-plane crystallographic orientation of the CrSe films on the Al$_2$O$_3$(001) and YSZ(111) substrates, we measured in-plane phi-scan of the (103) diffraction peak of the CrSe films. Figure 2(a) shows phi-scan of the XRD pattern for the CrSe(103) and Al$_2$O$_3$(104) showing 18 diffraction peaks of CrSe(103) with different intensities despite that CrSe has six-fold symmetry around the *c* axis. Around the weak intensity with a broad width (full-width at half-maximum Δphi ~ 4.3º) of CrSe(103) at phi = 0º, corresponding to the epitaxial relationship of CrSe(001)[100]//Al$_2$O$_3$(001)[100], there are two peaks with a stronger intensity with a sharp width (Δphi ~ 1.3º) at phi = ±19.1º. Here, phi is defined as an azimuthal angle with respect to the diffraction peak from Al$_2$O$_3$(104). The three peaks at phi = 0º and ±19.1º of CrSe(103) repeat every 60º along phi. These results indicate that growth of the in-plane ±19.1º-twisted domains are preferable compared to the aligned domain. The presence of twisted domains of CrSe is in stark contrast to the CrTe thin films on the Al$_2$O$_3$(001) substrate, where only the aligned domain was observed [4]. In contrast to the mixture of the in-plane aligned and ±19.1º-twisted domains on the Al$_2$O$_3$ substrate, the six-fold XRD pattern is clearly observed on the YSZ(111) substrate as shown in Fig. 2(b). The six-fold symmetry of phi-scan XRD is direct evidence of single-domain growth of CrSe with the epitaxial relationship of CrSe(001)[100]//YSZ(111)[1$\bar{1}$0].

    To interpret this contrast in the in-plane domain formation on the Al$_2$O$_3$(001) and YSZ(111) substrates, we discuss the two-dimensional (2D) atomic arrangement on the

surface of each substrate as shown in Fig. 2(c) and 2(d), respectively. First, we consider 3×4 unit-cells (UC) of CrSe rotated by +19.1° (navy solid lines) or that rotated by −19.1° (not shown) on the Al$_2$O$_3$ substrate in Fig. 2(c). A rhombus is formed by the four fourth nearest neighbor Cr atoms (indicated by black dotted circles) with their distance of 9.747 Å, which we found closely matches twice of the *a*-axis length of Al$_2$O$_3$ ($2a_{Al_2O_3}$ = 9.5178 Å). In this rotationally commensurate relation, the calculated mismatch is greatly reduced to +2.4% from the case of aligned epitaxial relationship (–22.6%). In this intuitive model, we took into account only four specific locations where the Cr atoms lie directly on the O atoms. However, the actual situation is more complicated because there should be other Cr and Se atoms with their sites not being located on the underlying Al and O atoms. Figure 2(d) shows the aligned atomic arrangement of Cr and Se on the YSZ(111) plane with the epitaxial relationship of CrSe(001)[100]//YSZ(111)[1$\bar{1}$0] determined by XRD pattern shown in Fig. 2(b). The Cr-Cr distance of 3.684 Å on (001) plane of CrSe almost matches the O-O distance on (111) plane of YSZ substrate, as discussed in Fig. 1(b). This relatively small lattice mismatch drives the single-domain formation on YSZ(111) substrate.

To characterize stability of rotational commensurate epitaxy on the Al$_2$O$_3$ substrate in a more realistic situation, we have performed the modeling of atomic arrangements through the DFT calculations. We choose two specific conditions, which correspond to the experimentally-observed 19.1°-twisted and aligned CrSe domains on the $Al_2O_3$(001) surface. Figures 3(a) and 3(b) show the side view of supercell of the 19.1°-twisted and aligned CrSe on the $Al_2O_3$ substrate, respectively, after reconstruction. Here, the Cr-terminated CrSe is placed on the O-terminated $Al_2O_3$ substrate in each alignment as the pristine condition. For both configurations studied, the first Cr layer from the interface (indicated by the blue arrow) are largely displaced in both directions parallel and perpendicular to the interface while the reconstruction of Cr atoms in the second layer (gray arrow) and above, and the displacement of the Se frameworks are found to be not significant. In addition, the surface O atoms on the $Al_2O_3$ substrate are also not much displaced. The large displacement of Cr atoms in the first layer may be a signature of formation of six-coordinate bonds with neighboring O and Se atoms.

Figure 3(c) shows the top-view supercell simulating the 19.1°-twisted domain (defined as yellow dashed line) on $Al_2O_3$ after reconstruction, where only the Cr atoms in the first and second layers as well as the first $Al_2O_3$ dioctahedral sheet are made visible for clarity. As can be seen, the Cr atoms at vertexes of the supercell in the first layer form the lattice-matched rhombus on the 2×2 UC of $Al_2O_3$ as indicated by the yellow dashed line, which is consistent with the intuitive speculation in Fig. 2(c). Inside the rhombus, three Cr atoms are displaced from the pristine position (see the almost undisplaced second Cr layer as a reference) to the nearest unoccupied octahedral sites of the $Al_2O_3$ dioctahedral sheet, while the other three Cr atoms share one unoccupied octahedral site. It turns out that (1) the Cr atom at the first layer forms an octahedron with three coordination to the O atoms on the $Al_2O_3$ surface (indicated by blue triangles) and the other three to Se atoms in CrSe, and (2) the location of the Cr site within the $CrSe_3O_3$ octahedron can be moved depending on the circumambient Al sites. Figure 3(d) shows the top-view supercell of aligned $CrSe/Al_2O_3$ interface. Similarly to the case of 19.1°-twisted domain, the Cr atom at vertexes of the 4×4 UC CrSe supercell in the first layer (yellow dashed line) is closely lattice-matched with 3×3 UC of $Al_2O_3$ and the Cr atoms in the first layer forms three coordination to O atoms on the $Al_2O_3(001)$ surface. We compare the stability of these two slab systems by calculating $E[\text{CrSe}] = (E[\text{total}] - E[Al_2O_3])/N_{\text{CrSe}}$, where $E[\text{total}]$, $E[Al_2O_3]$, and $N_{\text{CrSe}}$ are the total energy of the whole system, that for isolate slab $Al_2O_3$, and the number of CrSe included in the cell, respectively. Here, $E[\text{CrSe}]$ represents the energy of CrSe stacked on $Al_2O_3$ substrate. We found that $E[\text{CrSe}]$ calculated for the 19.1°-twisted domain is smaller than that for the aligned domain by 60 meV. This subtle energy difference between the two configurations is thermodynamically consistent with experimental observation of mixture of the 19.1°-twisted and aligned domains. The energy scale of the thin-film growth temperature at 750°C is $k_B T_{sub} \sim 65$ meV ($k_B$: the Boltzmann constant), which is comparable to this energy difference.

Finally, we present the impact of the single-domain formation on electrical transport property of the CrSe thin films. Figures 4(a) and 4(b) show temperature dependence of longitudinal resistivity $\rho_{xx}(T)$ and its derivative $d\rho_{xx}(T)/dT$ for the CrSe thin films on the Al$_2$O$_3$(001) (black solid line) and YSZ(111) (red) substrates. Both films exhibit a metallic $\rho_{xx}(T)$ behavior in whole temperature range down to 4.2 K being consistent with our previous report of the stoichiometric CrSe thin film [7]. In addition, the kink in $\rho_{xx}(T)$, which we ascribed to antiferromagnetic transition temperature [7], is commonly observed around $T_{kink} \sim 180$ K for both samples. This kink is more clearly observed as the development of $d\rho_{xx}(T)/dT$ in Fig. 4(b), indicating the comparable value of $T_{kink}$. The observed difference between the CrSe thin films on the Al$_2$O$_3$(001) and YSZ(111) substrates is the residual resistivity ratio (RRR) defined as $\rho_{xx}(T = 300$ K$)/\rho_{xx}(T = 4.2$ K$)$. Although the temperature dependence of carrier mobility is often used for characterizing scattering mechanisms, the extraction of the carrier mobility and carrier concentration cannot be applied to CrSe owing to the multi-carrier feature and a certain contribution of anomalous Hall resistivity [7]. Therefore, we restrict our discussion to the RRR value. The CrSe thin film on YSZ(111) exhibits RRR of 1.65, which is slightly larger than that on Al$_2$O$_3$(001) (RRR = 1.45). The superior metallic behavior, especially below $T_{\text{kink}}$, for the CrSe thin film on the YSZ(111) substrate with the suppressed twisted domains infers that the resistivity is partly governed by grain boundary scattering.

**IV. SUMMARY AND OUTLOOK**

In conclusion, we have investigated in-plane epitaxial relationship in thin-film growth of CrSe on the Al$_2$O$_3$(001) and lattice-matched YSZ(111) substrates. On the Al$_2$O$_3$ substrate, we observe formation of ±19.1°-twisted domains with respect to [100] of Al$_2$O$_3$ in the plane. The in-plan phi-scan of XRD pattern reveals that the twisted domain is dominant to the aligned domains with a large lattice mismatch of −22.6%. This specific twist angle can be considered by the rotational commensurate growth of CrSe on the Al$_2$O$_3$(001) surface. The 3×4 unit-cells of CrSe(001) plane rotated by ±19.1° has a small lattice mismatch of +2.4% on the 2×2 unit-cells of Al$_2$O$_3$. This estimation is further supported by DFT calculations implying that the rotational commensurate atomic arrangement is more energetically stable than the aligned configuration by 60 meV. On the YSZ(111) substrate, we demonstrate single-domain growth of the aligned CrSe thin film. The single-domain CrSe thin film exhibits in superior metallic conductivity to that on Al$_2$O$_3$ substate. Our demonstration of control of twisted-domain and single domain by selection of substrates pave the way toward further functionalizing Cr$X$-based thin films and their heterostructures.

**Figure caption**

**FIG. 1** (a) Schematic crystal structure of CrSe. (b) The *a*-axis length of Cr*X* compounds (*X* = Se and Te) and the O-O distance on YSZ(111) and $Al_2O_3$(001) substrates. (c)(d) A wide range of out-of-plane 2theta-omega scan of x-ray diffraction (XRD) patterns for the CrSe thin films deposited on (c) $Al_2O_3$(001) and (d) YSZ(111) substrates. (e)(f) The 2theta-omega scan of XRD pattern around CrSe(001) for the CrSe thin films deposited on (e) $Al_2O_3$(001) and (f) YSZ(111) substrates.

**FIG. 2** (a) In-plane azimuthal rotation (phi-scan) of x-ray diffraction (XRD) of (top) $Al_2O_3$(104) and (bottom) CrSe(103) for the CrSe thin film deposited on the $Al_2O_3$(001) substrate. (b) In-plane phi-scan of XRD of (top) YSZ(400) and (bottom) CrSe(103) for the CrSe thin film deposited on the YSZ(111) substrate. (c) Two-dimensional (2D) atomic arrangement at interface of CrSe with $Al_2O_3$. The 3×4 unit-cells of CrSe is rotated in the plane by 19.1° in counter-clockwise. Blue shaded area represents (001) plane of the $Al_2O_3$ unit-cell (UC). At the location marked by black dashed circles, the Cr atom forms lattice-matched rhombus on 2×2 $Al_2O_3$. (d) 2D atomic arrangement of CrSe on the (111) surface of the lattice-matched YSZ substrate with the epitaxial relationship of CrSe(001)[100]//YSZ(111)[1$\bar{1}$0]. Orange shaded area represents the (111) plane of the

YSZ UC.

**FIG. 3** (a)(b) Side-view supercell of (a) 19.1°-twisted and (b) aligned CrSe on the Al$_2$O$_3$(001) substrate. The Cr, Se, Al, and O atoms are represented by blue, green, light blue, and red spheres, respectively. Blue and gray arrows indicate the first and second Cr layers from the CrSe/Al$_2$O$_3$ interface, respectively. Top-view supercell of (c) 19.1°-twisted and (d) aligned CrSe on the Al$_2$O$_3$ substrate. For clarity, only Cr atoms in the first layer (blue) and those in the second layer (gray) are shown with the top Al$_2$O$_3$ layer. Yellow dashed lines indicate boundary of supercells. These figures are drawn by VESTA [34].

**FIG. 4** Temperature dependence of (a) longitudinal resistivity $\rho_{xx}(T)$ and (b) its derivative d$\rho_{xx}$/d$T$ for the CrSe thin films grown on Al$_2$O$_3$ (black solid line) and YSZ (red solid line) substrates.

**Acknowledgment**

This work was supported by JST, PRESTO Grant No. JPMJPR21A8, Japan and JSPS KAKENHI Grant Nos. JP23H01686, JP22H01182, JP23K22453, and JP24K21531, Iketani Science and Technology Foundation, and Tanikawa Foundation.

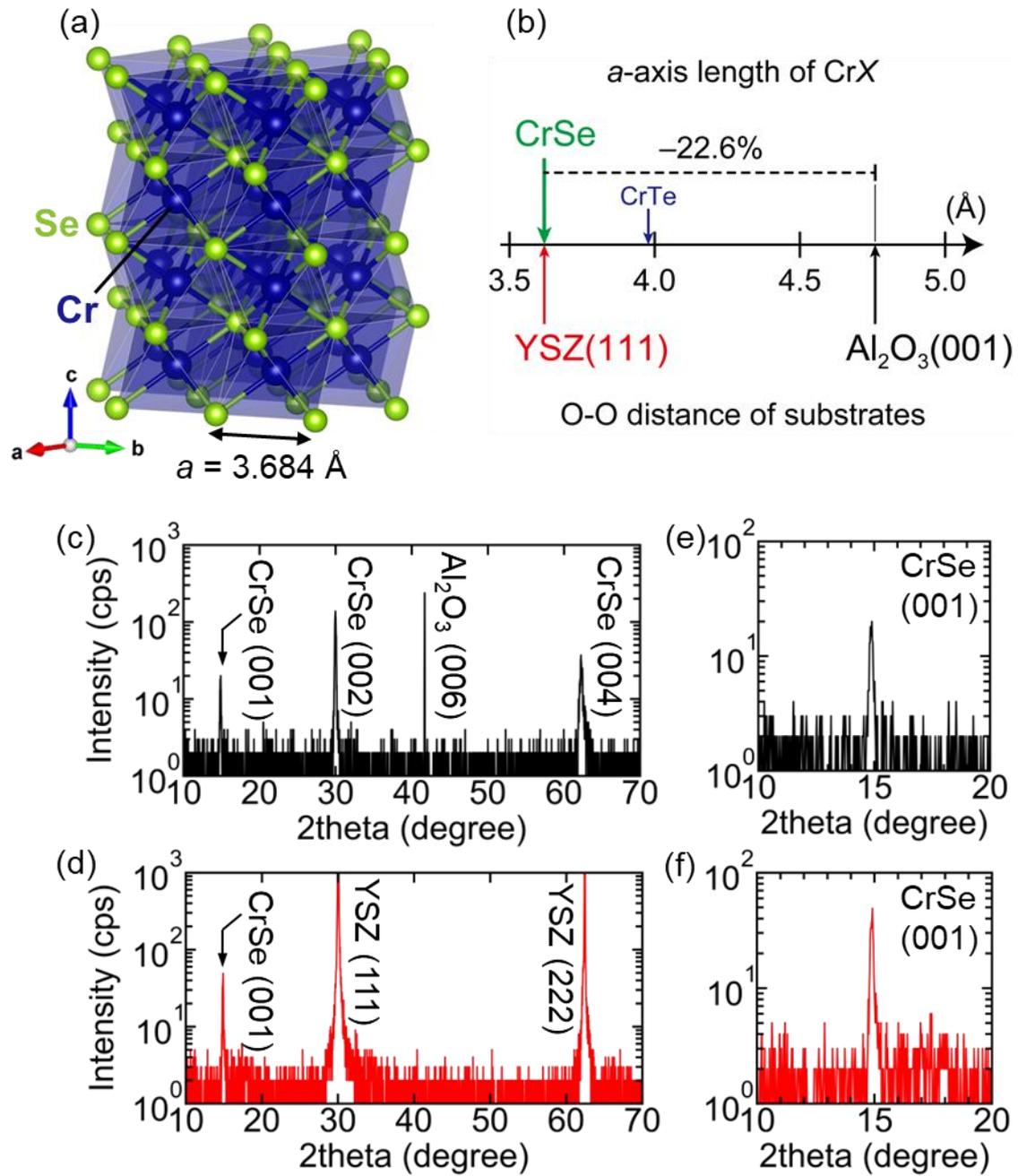

FIG. 1 (double column) Y. Tajima *et al*.

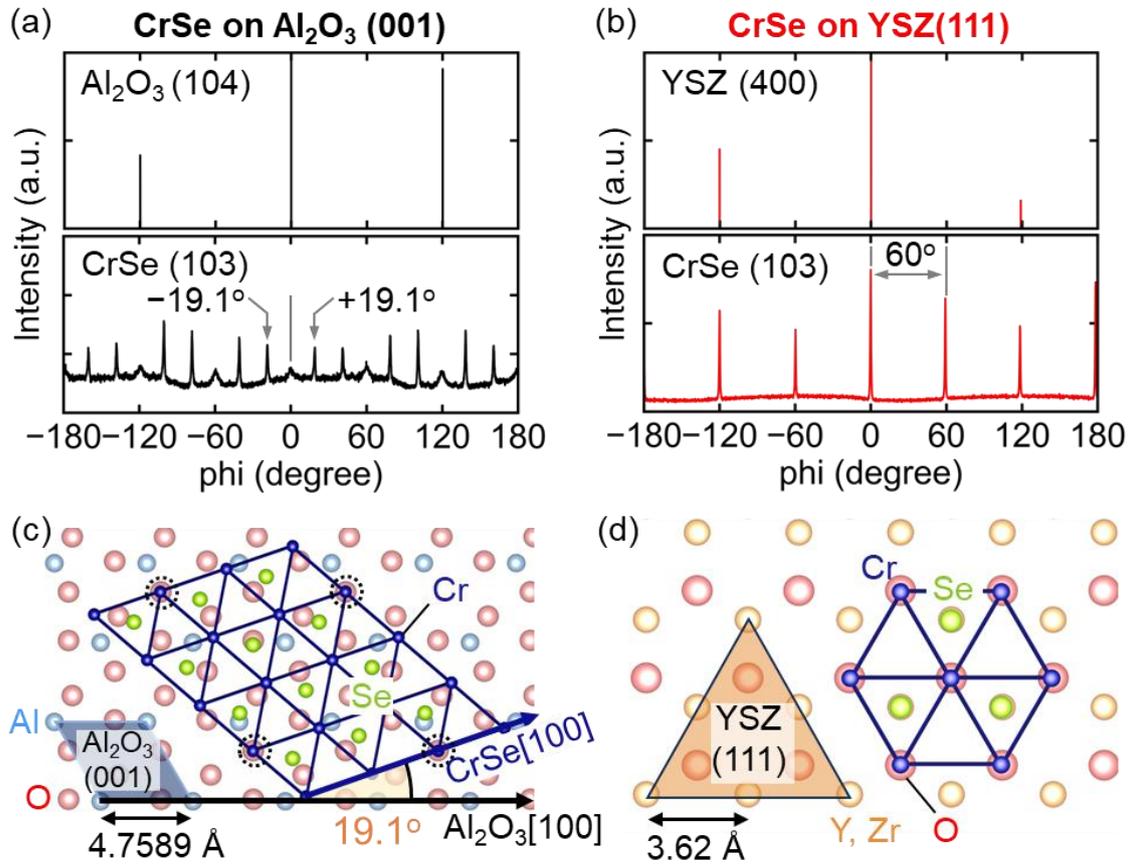

FIG. 2 (double column) Y. Tajima *et al*.

## 19.1°-twisted CrSe on Al$_2$O$_3$

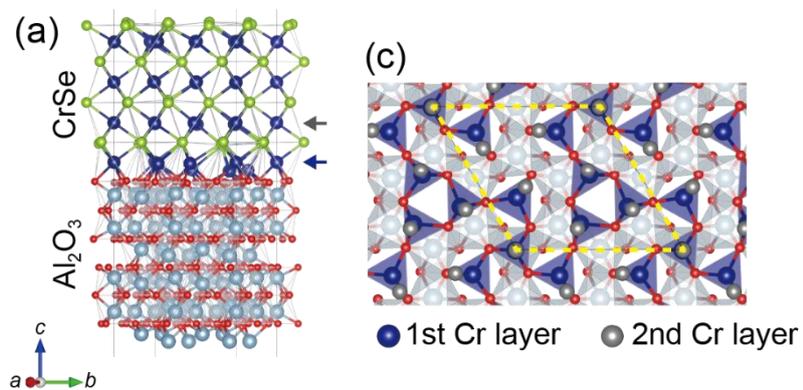

## Aligned CrSe on Al$_2$O$_3$

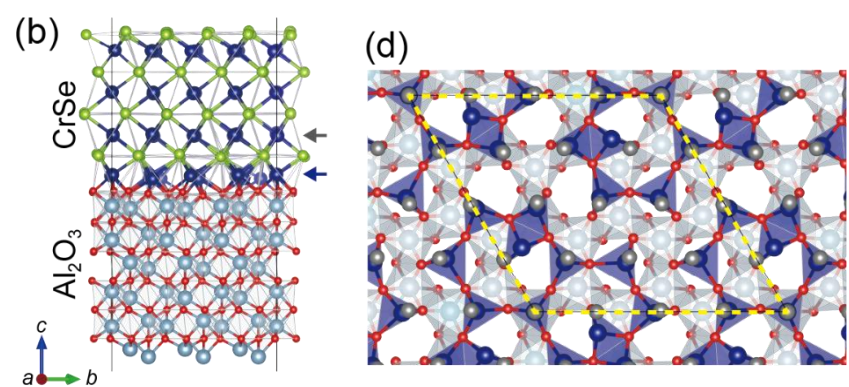

FIG. 3 (single column) Y. Tajima *et al*.

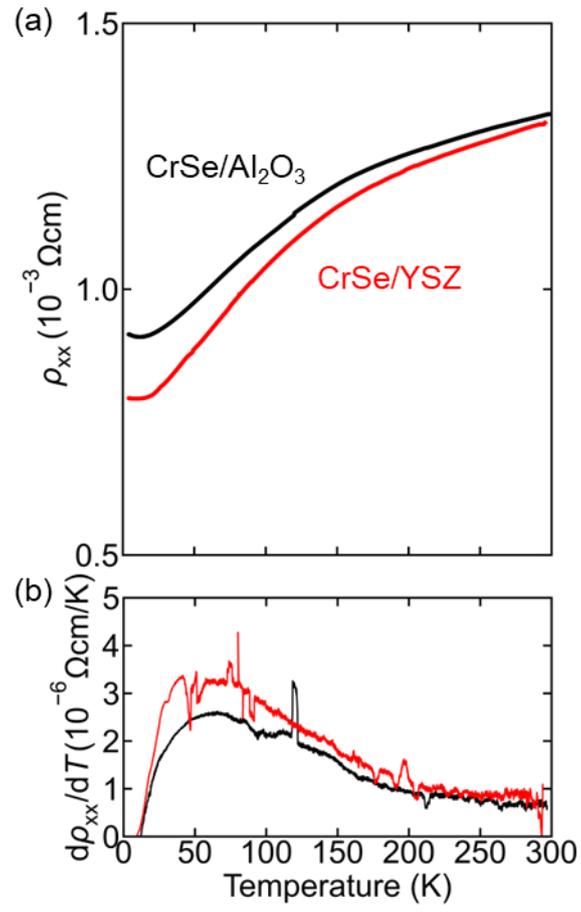

FIG. 4 (single column) Y. Tajima *et al*.